# Electroneutrality Breakdown and Specific Ion Effects in Nanoconfined Aqueous Electrolytes Observed by NMR

Zhi-Xiang Luo[1], Yun-Zhao Xing[2], Yan-Chun Ling[2], Alfred Kleinhammes[1], Yue Wu[1,2*]

[1]Department of Physics and Astronomy, University of North Carolina, Chapel Hill, NC 27599-3255, USA

[2]Department of Applied Physical Sciences, University of North Carolina, Chapel Hill, NC 27599-3216, USA

**Abstract**

Ion distribution in aqueous electrolytes near the interface plays critical roles in electrochemical, biological and colloidal systems and is expected to be particularly significant inside nanoconfined regions. Electroneutrality of the total charge inside nanoconfined regions is commonly assumed *a priori* in solving ion distribution of aqueous electrolytes nanoconfined by uncharged hydrophobic surfaces with no direct experimental validation. Here, we use a quantitative nuclear magnetic resonance approach to investigate the properties of aqueous electrolytes nanoconfined in graphitic-like nanoporous carbon. Substantial electroneutrality breakdown in nanoconfined regions and very asymmetric responses of cations and anions to the charging of nanoconfining surfaces are observed. The electroneutrality breakdown is shown to depend strongly on the propensity of anions toward the water-carbon interface and such ion-specific response follows generally the anion ranking of the Hofmeister series. The experimental observations are further supported by numerical evaluation using the generalized Poisson-Boltzmann equation.

Electric double layer (EDL) near the interface is of fundamental importance in various applications ranging from redox reactions in electrochemistry to colloidal particles assembly[1] and DNA sequencing[2]. The neutrality of the total charge is an important condition in deriving the ion distribution near the interface in the EDL theory. For an uncharged hydrophobic surface such as the water/air interface, positive and negative ions can still be separated in the interfacial region (~10 Å) due to different propensities toward the interface between cations and anions[3, 4, 5, 6]; such effect is called specific-ion effect[7, 8, 9, 10, 11] since it is driven by nonelectrostatic interactions that varies significantly between different ions even for ions with the same electrovalency (*e.g.*, $F^-$ and $I^-$). In the scenario of aqueous electrolytes confined by hydrophobic surfaces where the pore size is comparable in size to the interfacial region determined by the specific-ion effect, a natural question raised is how the tendency of charge separation near the interface reconciles with electroneutrality inside nanoconfined regions. Could electroneutrality of the total charge in fact be violated substantially inside nanoconfined regions driven by the specific-ion effect? Theoretical studies nearly always take the total charge neutrality inside nanoconfined regions for granted and experimental evaluation of electroneutrality inside nanoconfined regions is lacking. Such evaluation could contribute significantly to our understandings of some very important processes such as energy storage in supercapacitor[12], ion transport through nanochannels[13], and ionic processes in proteins[7].

Nanoporous carbon with graphitic-like internal surfaces provides an ideal model system for investigating the electroneutrality in nanoconfined aqueous electrolytes using nuclear magnetic resonance (NMR). Previous studies showed that fluid inside carbon nanopores exhibits a different NMR chemical shift from that outside the nanopores due to the ring current effect, which gives rise to a nucleus independent chemical shift (NICS)[14, 15, 16, 17, 18]. This shift provides a clear NMR marker for selectively and quantitatively monitoring the electrolyte inside nanometer-sized regions confined by hydrophobic graphitic-like carbon surfaces. It provides an excellent tool for determining quantitatively the cation and anion concentrations inside nanopores. Ions, especially anions, can be ordered by their influence on a vast variety of specific ion effects, called the Hofmeister series[7, 10, 19]. A typical ranking is $SO_4^{2-} < F^- < Cl^- < Br^- < NO_3^- < I^- < BF_4^- < ClO_4^-$ for some anions with increasing protein solubility in aqueous electrolytes to the right side (often referred to as the chaotropic side)[10]. Evaluating the electroneutrality with systematic change of anions according to the Hofmeister series provides another avenue for revealing the potential electroneutrality breakdown caused by the ion-specific interfacial effect. Here we report such a quantitative NMR study of the ion concentrations in nanoconfined aqueous electrolytes. Hydrophobic graphitic-like porous carbon is used as a model system to provide the nanoconfinement. Direct experimental evidence is observed for a significant electroneutrality breakdown of the total charge inside nanometer-sized regions even when the carbon material is uncharged. Interfacial specific ion effects and ion-ion correlations are shown to play crucial roles in determining the degree of electroneutrality breakdown inside nanopores. The importance of the specific-ion interfacial effect is further revealed by the asymmetric and nonlinear responses of cation and anion concentrations to the external charging of the nanoconfining carbon walls. Such information was obtained using a charge-controlling device built into the NMR probe. The experimental results are further validated by a numerical calculation using the generalized Poisson-Boltzmann (PB) equation in nanopores, demonstrating that specific-ion interfacial effect can indeed dominate the electrostatic interactions leading to the breakdown of electroneutrality inside nanoconfined regions.

## Results

**Electroneutrality breakdown in nanoconfinement**

A high quality nanoporous carbon derived from polymer poly(etheretherketone) (PEEK)[20, 21] is used to provide the hydrophobic nanoconfinement in this work (See Methods). The activated carbon sample is designated as P-40 and the average pore size is 0.9 nm (wall surface to wall surface assuming a slit-shaped pore) and 1.2 nm (carbon to carbon centers) according to the previous study[15]. Unless specified, all results discussed here refer to that obtained using P-40. However, activated carbon with pore size of 1.9 nm (carbon to carbon centers), labeled P-92, was also used in the current study and will be mentioned as well. The capability of NMR approach to selectively and quantitatively study nanoconfined fluids is demonstrated in Figure 1a where the $^1$H, $^{19}$F, and $^{23}$Na static NMR spectra of NaBF$_4$ electrolyte injected into P-40 are shown (See Methods). All spectra consist of two peaks. The peak centered at 0 ppm, chosen as the reference, comes from electrolyte outside the nanopores while the peak centered at -7 ppm is from electrolyte inside nanopores[22]. All three nuclei show the same chemical shift at -7 ppm because the shift is completely determined by the NICS effect.

Since the NMR signal is proportional to the number of spins, numbers of cations outside and inside the nanopores can be determined by the $^{23}$Na peak intensities at 0 and -7 ppm, respectively. Similarly, numbers of BF$_4^-$ anions and water molecules outside and inside nanopores can be determined from the corresponding peak intensities of the $^{19}$F and $^1$H NMR spectra, respectively. From these numbers the cation and anion average concentrations inside nanopores can be determined (See Methods). Figure 1b shows the normalized ion concentrations, $c/c_0$, where $c$ is the average ion concentration in nanopores and $c_0$ is the injected electrolyte concentration (1 mol/kg except for NaF 0.8 mol/kg due to its lower water solubility), for NaF, $NaNO_3$, $NaBF_4$ electrolytes in P-40 and $NaBF_4$ electrolyte in P-92. One of the surprising phenomena revealed by measurements shown in Figure 1b is the drastic concentration difference between cations and anions, particularly significant in nanoconfined aqueous electrolytes of $NaNO_3$ and $NaBF_4$. The concentration inside nanopores is $c/c_0$=1.92 for $BF_4^-$ and $c/c_0$=0.64 for $Na^+$. In the larger pore P-92 sample, the concentration inside nanopores is $c/c_0$=1.34 for $BF_4^-$ and $c/c_0$=0.70 for $Na^+$. The anomalous concentration difference is a strong indication of the electroneutrality breakdown of the total charge inside the nanopores. As expected, the electroneutrality breakdown is less in the larger pore P-92 sample but it is nevertheless still very significant.

The possibility that the electrolyte neutrality might be maintained by other ions such as $H^+$, $OH^-$ or trace impurities can be ruled out in the current experimental approach. Take NaF electrolyte in P-40 as an example to estimate the amount of $H^+$ and $OH^-$. The PEEK derived activated carbon is of high quality and contains very few surface functional groups[21, 23] that does not produce $H^+$ or $OH^-$. So we can conclude that all the $H^+$ and $OH^-$ in this system are from water dissociation (depending on the point of zero charge and pH, the activated carbon can be positively or negatively charged, but the source of the charge still comes from water dissociation). Since only limited electrolyte is injected into the activated carbon, the electrolyte amount in the intergranular space is only about three times that inside carbon nanopores. The intergranular electrolyte pH is measured to be 10 in the slurry. Therefore the net charge due to $H^+$ or $OH^-$ inside carbon nanopores is at most $3 \times 10^{-4}$ mol/kg which is negligible compared to the ion concentration inside nanopores ($Na^+$ 0.17 mol/kg, $F^-$ 0.24 mol/kg). Similar estimate can be applied to other ions and the trace impurities (less than 1%) in the as-purchased chemicals.

This shows that the electroneutrality breakdown of the total charge inside carbon nanopores is an intrinsic property of nanoconfined aqueous electrolytes in this system.

**Specific ion effects on ion concentrations**

Another intriguing phenomenon beyond the electroneutrality breakdown revealed by the data in Figure 1b is the strong influence of anions on the $Na^+$ concentration. Although the experiments are carried out with similar electrolyte concentrations and electrolyte/carbon ratios, the $Na^+$ concentrations vary significantly among different electrolytes. $Na^+$ concentration for NaF electrolyte in nanopores is highly suppressed while that for $NaNO_3$ is very close to the injected electrolyte concentration. It is interesting to note that the anion concentration increases in the order $F^- < NO_3^- < BF_4^-$ with $F^-$ concentration being also highly suppressed in the nanopores while $NO_3^-$ and $BF_4^-$ concentrations being greatly enhanced. The $F^- < NO_3^- < BF_4^-$ ranking based on their concentrations is fully consistent with the ranking of the Hofmeister series where the anions are known to have different propensities for a hydrophobic surface[7].

Systematic testing on a series of sodium salt electrolytes whose anions are chosen from the Hofmeister series $SO_4^{2-} < F^- < Cl^- < Br^- < NO_3^- < I^- < BF_4^- < ClO_4^-$ provides more insights into the anion-dependent $Na^+$ concentrations inside nanopores. The normalized average $Na^+$ cation concentration $c/c_0$ for the sodium salt series is shown in Figure 1c. $Na^+$ concentration inside nanopores increases gradually from $Na_2SO_4$ to $NaClO_4$ following the anion Hofmeister series with $NaNO_3$ being a clear exception (and slightly for NaI). It is of note that $Na^+$ concentration inside nanopores is highly suppressed to $c/c_0 <0.2$ for $Na_2SO_4$ and NaF, <0.4 for NaCl and NaBr, and <0.7 for NaI and $NaBF_4$. Even though $I^-$ and $BF_4^-$ ranked to the right side (the chaotropic side) of $NO_3^-$ in the Hofmeister series, $c/c_0$ =0.86 for $NaNO_3$ is significantly higher than that of NaI and NaBF$_4$. It is also of note that unlike other electrolytes, Na$^+$ concentration for NaClO$_4$ in nanopores is substantially enhanced ($c/c_0$=1.32) rather than suppressed compared to that of the bulk concentration. Because limited amount of electrolyte is added to the sample, Na$^+$ concentration outside the nanopores also differs from $c_0$. The Na$^+$ concentration in nanopores normalized by that outside nanopores shows slightly different values from $c/c_0$ but maintains the same trend of Na$^+$ concentration increase including the NaNO$_3$ anomaly.

The strongly anion-dependent Na$^+$ concentration inside carbon nanopores revealed by the quantitative NMR analysis demonstrates the intriguing interplay between cations and anions. Na$^+$ is a strongly hydrated cation with hydration free energy of -87 kcal/mol, hydration number of 5 to 6 in the first hydration shell[24, 25], and no propensity for the interface[8]. In fact, strong hydration leads to a free energy barrier of several $k_B T$ ($T$=300 K) or higher for Na$^+$ ions to enter the hydrophobic nanopore with diameter less than 2 nm[26]. This is clearly reflected by the low value of $c/c_0$<0.2 for Na$^+$ in NaF. Theory predicts F$^-$ < Cl$^-$ < I$^-$ to be the ranking based on their propensity for the interface[8]. This trend is expected to hold for most anions in the Hofmeister series where the hydration enthalpy becomes less negative toward the chaotropic side of the series[27]. The differences among those anions give rise to the different Na$^+$ cation concentrations.

**Numerical calculation**

More insight can be gained by looking at the various factors determining the ion distribution near the interface. The ion distribution for ion $i$ with valency $z_i$ is given by[28]

$$\rho_i(x) = \frac{\exp\left(\frac{\mu_i}{k_B T}\right)}{\Lambda_i^3} \exp\left(-\frac{z_i e\psi(x) + V_i^{ext}(x) + corr_i(x)}{k_B T}\right) \quad (1)$$

where $e$ is the elementary charge, $\Lambda_i$ is the de Broglie thermal wavelength of ion $i$, $\mu_i$ is the chemical potential of ion $i$, $\psi(x)$ is the local electrostatic potential at the location $x$ inside nanopores, $V_i^{ext}(x)$ is the ion-surface potential that depends on the ion-specific propensity for the interface[28, 29], and $corr_i(x)$ is the free energy contribution from ion-ion correlations, a quantity that requires molecular scale structural information to obtain such as via theory and molecular dynamics simulations[28]. $corr_i(x)$ depends on both the ion-specific short-ranged pair potential and the counterion concentration, which is implicitly affected by the electrostatic potential $\psi(x)$. The ion concentration measured by NMR is the averaged value over the pore width $d$: $\bar{\rho}_i = \frac{1}{d}\int_0^d \rho_i(x)dx$. Although ion-surface potential $V_i^{ext}(x)$ is generally position dependent and has an oscillatory character[29, 30], it is expected that the effective potential $\bar{V}_i^{ext}$ for anions, defined by $(-\beta\bar{V}_i^{ext}) = \frac{1}{d}\int_0^d \exp(-\beta\bar{V}_i^{ext}(x))dx$, ranks according to the Hofmeister series in P-40 nanopores. As such, a larger $-\bar{V}_i^{ext}$ value for the more chaotropic anion would lead to a higher anion concentration and that would attract more Na$^+$ counterions into nanopores electrostatically. Of course, this argument does not take into considerations of the ion-ion correlations (i.e. $c_i(x) = 0$).

Numerical calculation of the generalized PB equation[7] in a slit-shaped nanopore is carried out to reveal the mechanism of the electroneutrality breakdown in nanoconfined aqueous electrolytes (See Methods). Since the ion-surface potential[29, 31] and ion-ion correlation functions[30, 32, 33] from MD simulation are not available for our system, we ignore the ion-ion correlations at this moment and use a simplified ion-surface interactions potential[34] $V_i^{ext}(r) = \frac{B_i}{r^3}$.

Here $r$ is the distance from the ion to the confining surface; $B_i$ characterizes the ion-surface interaction strength whose value is about few $k_B T$ near the surface[35]. Because the boundary condition on the metal plate is unknown (even though the net charge on the metal plate is zero, we could not assume the surface charge is zero because the inner surface and outer surface may carry induced charges of an equal amount but opposite signs), electrostatic potentials both inside and outside the nanopore need to be solved jointly in order to find the ion distribution inside nanopores (See Methods).

Ion distribution in a 1 nm pore is illustrated in Figure 2b. In this calculation we use $B_i$ value $-58 \times 10^{-50}$ Jm$^3$ and $46 \times 10^{-50}$ Jm$^3$ (about $5 k_B T$ at 0.3 nm from the surface) for anions and cations, respectively. Anions are preferentially adsorbed on the surface because an attractive potential $V_i^{ext}(x)$ is chosen whereas cations are repelled from the surface. The average anion concentration in the nanopore is much higher than that of cation, indicating an electroneutrality breakdown of the total nanoconfined charge. Clearly, the ion-surface interaction is responsible for such electroneutrality breakdown. If only Columbic interaction is considered as in the Gouy-Chapman theory, i.e. $V_i^{ext}(x) = 0$, bulk concentration will be obtained in nanopores and the total charge is neutral.

The ion concentrations depend on the pore size and the strength of the anion-surface interactions. The average ion concentration versus pore size is shown in Figure 2c. The $B_i$ values are the same as in Figure 2b. The electroneutrality breakdown is prominent only when the pore size is less than 2 nm. As the pore size increases, the concentration difference between cations and anions disappears and both ion concentrations approach that of the bulk value. Figure 2d shows the average ion concentration versus $B_-$, demonstrating the specific ion effects on the extent of the electroneutrality breakdown in 1 nm pores. Here, $B_+$ is fixed at $46 \times 10^{-50}$ Jm$^3$ while $B_-$ varies from $40 \times 10^{-50}$ Jm$^3$ to $-70 \times 10^{-50}$ Jm$^3$ to represent the increased ion propensity for the interface. The average anion concentration increases as expected when $B_-$ becomes more negative. Although $B_+$ is kept unchanged, cation concentration also increases because of the increased electrostatic attraction to anions. The electroneutrality breakdown is more prominent as the propensity difference between cations and anions grows. It is of note that the numerical calculation here shows a monotonic increase of the cation concentration which could not explain the anomaly of high Na$^+$ concentration in nanoconfined NaNO$_3$ electrolyte. This is because ion-ion correlations are not included in this calculation.

The ion-ion correlations based on electrostatic and ion-specific interactions are predicted to be of crucial importance in nanoconfined electrolytes[28, 30, 32, 36, 37]. Although the preferentially adsorbed anions in the nanopores could attract Na$^+$ cations via electrostatic interactions as demonstrated by both the experiments and the simulation, the higher Na$^+$ concentration associated with NaNO$_3$ electrolyte is not due to the anomalous interfacial affinity of NO$_3^-$ since its concentration is consistent with the ranking of the Hofmeister series, i.e. lower than the BF$_4^-$ concentration (Figure 1b). Clearly, specific ion-ion correlations must be invoked to explain the abnormal Na$^+$ concentration in NaNO$_3$. Correlations of Na$^+$ with NO$_3^-$ appear to be stronger than that with I$^-$ and BF$_4^-$, suggesting a more negative effective correlation $\overline{corr_+}$, defined by $\exp(-\beta \overline{corr_+}) = \frac{1}{d}\int_0^d \exp(-\beta corr_+(x))dx$, for Na$^+$ inside the nanopores. It is interesting to note that the formation of solvent-separated Na$^+$ and NO$_3^-$ ion pairs in bulk electrolyte has been recognized by both computational and experimental studies[38, 39]. The formation of solvent-separated Na$^+$ and ClO$_4^-$ ion pairs was also found in bulk electrolyte[40]. Such molecular scale ion-ion correlations could become more significant at the interface and in nanoconfined environment giving rise to the observed anomaly in the Na$^+$ concentration of NaNO$_3$ and the substantially enhanced Na$^+$ concentration in NaClO$_4$ aqueous electrolyte.

**Ion concentrations in charged nanopores**

To further demonstrate how the non-electrostatic specific ion effects including ion-ion correlations dominate the electrostatic interactions inside the nanopores and lead to the intriguing electroneutrality breakdown, ion concentrations versus confining wall surface charging is measured with the *in-situ* NMR[41, 42, 43]. As a model system to investigate electrolyte properties under hydrophobic nanoconfinement, the electric conducting property of activated carbon is an additional benefit which allows fine control of the surface charge to tune the electrostatic interactions. This is achieved by incorporating a device similar to a supercapacitor[44] into an NMR probe (See Methods). As illustrated in Figure 3a, two electrodes made of P-40 are separated by a glass fiber and immersed in an aqueous electrolyte. Voltage can be applied between the two electrodes to change the surface charge while NMR spectrum is acquired *in-situ*. The ion concentration inside the nanopores versus charging voltage is measured for a single electrode while the other one is covered with a copper foil to shield the radio frequency pulse and signal.

Figure 3b shows the ion concentrations inside P-40 nanopores versus charging voltage for NaBF$_4$ electrolyte measured by $^{19}$F and $^{23}$Na NMR. On positive charging (+V), both Na$^+$ and BF$_4^-$ concentrations respond linearly with the charging voltage. The influence of the non-electrostatic interactions is reflected on the huge initial concentration difference at 0 V. Because the surface is already crowded with anions at 0 V, further positive charging is unlikely to bring in more anions to the surface where non-electrostatic interactions are dominant. Therefore, such linear behavior is expected when the ion concentration change is mainly due to ions away from the interface and is affected by the change of electrostatic interactions[28]. In contrast, both $Na^+$ and $BF_4^-$ exhibit nonlinear behavior on negative charging (-V). $Na^+$ concentration increases with voltage from 0 to 0.6 V but then starts to decrease with further negative charging. Concomitantly, the initial linear decrease of $BF_4^-$ concentration levels off beyond 0.6 V. The nonlinear behavior, particularly the unexpected $Na^+$ concentration decrease with negative charging beyond 0.6 V, demonstrates the competing effect between the ion-ion correlations and the ion-surface electrostatic interactions. The attractive Coulomb interaction between $Na^+$ and the negatively charged surface tends to bring $Na^+$ into the nanopores whereas the decreased $BF_4^-$ concentration favors dragging $Na^+$ out of the nanopores. When the latter effect dominates, the $Na^+$ concentration can actually decrease with further negative charging as observed in Figure 3b. It is also interesting to note that even at 1.0 V charging, the $BF_4^-$ concentration in nanopores is still higher than that of $Na^+$, demonstrating the strong ion-surface attractions that overcomes the enormous Coulombic forces due to the net charge in the nanopores and the repulsion between anions and the negative charged surface.

The influence of anions on the cation's behavior via ion-ion correlations is evidenced by comparing $Na^+$ behaviors between $NaBF_4$ and $NaNO_3$ electrolytes shown in Figure 3c. For the convenience of comparison, the concentration has been normalized by their respective value at 0 V. On positive charging, $Na^+$ concentrations in both $NaBF_4$ and $NaNO_3$ decrease linearly because they are mainly affected by the change in electrostatic interactions. However, drastically different behaviors are observed on negative charging: while response of $Na^+$ in $NaBF_4$ electrolyte first increases then decreases, the $Na^+$ concentration for $NaNO_3$ almost does not change with charging voltage, indicating that the correlation between $Na^+$ and $NO_3^-$ is stronger than that in $NaBF_4$. The Coulombic attraction on the cations by the negatively charged surface is completely compensated by the ion-ion correlations which drags $Na^+$ out of the nanopores when the anions are repelled from the nanopores.

**Discussion**

In this study, quantitative NMR measurements and numerical simulations are employed to investigate the electroneutrality condition in nanoconfined aqueous electrolytes. Substantial electroneutrality breakdown of the total charge is observed inside uncharged activated carbon nanopores. The ion-specific interfacial interactions and ion-ion correlations are found to play critical roles in determining the extent of the electroneutrality breakdown. These effects are further investigated in charged carbon nanopores which lead to strong asymmetric responses of cations and anions to the confining wall surface charging. Moreover, anions impose great influence on the cation's behavior via ion-ion correlations.

Our study demonstrates that graphitic-like porous carbon provides an ideal model system and the novel *in-situ* NMR approach opens a new avenue for quantitative experimental evaluations of various ion-specific interactions near the interface and under nanoconfinement. Although our work is based on aqueous electrolytes, it can be generally applied to other systems such as organic electrolyte and ionic liquids where the strong ion-specific properties beyond their electrovalencies (e.g. ion solvation, interaction with the surface, ion-ion correlations) are also of relevance. The NMR approach is also of great value for validating theories[30, 45, 46] where the possibility of nanoconfinement-induced electrolyte non-neutrality in aqueous electrolytes is often ignored in computational studies which commonly assume *a priori* a neutrality of the total charge in nanoconfined regions. The findings revealed by the NMR study have broad implications because the electroneutrality breakdown in aqueous electrolytes can be very substantial in nanoconfined regions, which exist in many systems including proteins, desalination devices, colloidal suspensions and supercapacitors.

**Methods**

**Carbon material P-40 preparation**

Porous carbon used in this work is derived from high-temperature polymer poly(etheretherketone) (PEEK) using a procedure modified from previous reports[15, 47]. PEEK pellets are carbonized at 900 ℃ for 30 minutes in Argon atmosphere. The carbonized chunk is then cooled down to room temperature and subsequently ground into small particles of approximately 0.5 mm in diameter. The pulverized sample is activated at 900 ℃ under water vapor for a designated time to achieve certain burn-off percentage. The pulverization ensures uniform activation which leads to a narrow pore size distribution. The mass reduction achieved by the activation process is 40% and the corresponding sample is designated as P-40. The average pore size of P-40 is 0.9 nm (wall surface to wall surface assuming a slit-shaped pore) and 1.2 nm (carbon to carbon centers) according to the previous study[15]. The sample which has mass reduction 92% during the activation process is labeled as P-92, whose pore size is 1.9 nm (carbon to carbon centers).

**Nanoconfined electrolytes preparation**

The sodium salts are purchased from Sigma-Aldrich and used as purchased without further purification. The purity is >99.0% expect for $NaBF_4$ (>98%). The aqueous electrolytes are prepared to contain $Na^+$ cations 1 mol/kg except for NaF (0.8 mol/kg because of its lower solubility in water). A simple procedure is followed for preparing the nanoconfined aqueous electrolyte. In general, 30 mL electrolyte is injected into 20 mg P-40 sample. The mixture is then tightly sealed in an NMR sample tube. P-40 has a pore volume of 0.5 cm$^3$/g therefore 30 mL electrolyte is sufficient to fill the nanopore and about two thirds of the electrolyte is left in the intergranular space.

**Device to control carbon surface charging**

The device used to control carbon surface charging is comprised of two electrodes made of pure P-40 separated by a glass fiber and immersed in the aqueous electrolyte (1 mol/kg $NaBF_4$ or $NaNO_3$). Each electrode is 3 mm long and 2.5 mm in diameter. One electrode is shielded with a copper foil so that the detected NMR signal comes only from a single electrode. Potential is applied between the two electrodes. The charging principal is similar to a supercapacitor. In brief, cations are driven away from the surface and anions are attracted to the surface on positive charging such that the electric charge on the surface is maintained to balance the net charge in the liquid side.

**Static NMR on non-charged P-40**

$^1$H, $^{23}$Na, $^{19}$F (for NaF and $NaBF_4$), and $^{15}$N (for $NaNO_3$, $^{15}$N enriched) static spectra on the electrolyte/P-40 mixture are measured with a 400 MHz pulsed NMR system at 293 K. A single pulse is used for data acquisition and last delay is set long enough to make sure the signal is fully recovered after each scan. The 90 degree pulse of $Na^+$ inside P-40 nanopores was shown to be the same as $Na^+$ in the intergranular space as well as in pure aqueous electrolyte solution.

Furthermore, there are no sidebands under 7 kHz magic angle spinning. All these indicate that the quadrupole interaction effect is negligible for $^{23}$Na NMR.

**Ion concentration calculation**

The two peaks in the $^{23}$Na NMR spectrum (representing ions in the nanopores and ions in the intergranular space) are well separated and are deconvoluted to obtain the intensities $A_{in}$ (inside nanopores) and $A_{out}$ (outside nanopores). Since the total number of Na$^+$ cations $n_{tot}$ associated with the entire NMR spectrum is known based on the amount of the injected electrolyte, the portion inside P-40 nanopores could be calculated by $n_{in} = n_{tot} \frac{A_{in}}{A_{in}+A_{out}}$. Using the same procedure the amounts of water inside and outside nanopores can be determined. From these numbers we can calculate the Na$^+$ concentration $c$ inside P-40 nanopores. The concentrations of BF$_4^-$ and NO$_3^-$ inside and outside nanopores can be determined similarly.

***In-situ* NMR on charged P-40**

*In-situ* $^{19}$F, $^{23}$Na NMR experiment is carried out on a homemade probe which is equipped with a charging system controlled by Labview. The device to control P-40 surface charging is charged from 0 V to 1.0 V with a step of 0.1 V. Static NMR spectrum is acquired when the charging reaches equilibrium, typically after 2 hours. For $^{19}$F NMR, the last delay is 5 s and the spin-lattice relaxation time ($T_1$) is 0.7 s. For $^{23}$Na, the last delay is 0.5 s and $T_1$ is 20 ms. Charging has little effect on the $T_1$ relaxation times and the 90 degree pulses for both $^{19}$F and $^{23}$Na. Quadrupole interaction effect for $^{23}$Na is confirmed to be negligible.

**Numerical calculation**

The nanopore confinement model is illustrated in Fig. 3a. Two conducting metal plates are immersed in 1 mol/L 1:1 electrolyte to simulate slit-shaped carbon nanopore with pore size $d$. Water is assumed to be a continuum with a dielectric constant $\varepsilon = 78.5$. The closest distance $x_{min}$ that ion can approach the surface is limited by the finite ion size. Here we use the typical hydrated ion radii[48] 0.35 nm as $x_{min}$ for both cation and anion.

The electrostatic potential near the solid liquid interface is described by

$$\varepsilon\varepsilon_0 \frac{d^2}{dx^2}\psi(x) = -\sum_i z_i e \rho_i^0 \exp(-\frac{z_i e\psi(x) + V_i^{ext}(x)}{k_B T}) \quad (2)$$

Since the system is symmetric from the center of the pore, designated as $x=0$, we only need to consider the region $x>0$. In the bulk electrolyte side, the ions only experience surface interactions from one side thus $V_{i,bulk}^{ext}(x) = \frac{B_i}{(x - x_2 + x_{min})^3}$, whereas inside the nanopore, ions experience interactions from both surfaces such that $V_{i,pore}^{ext}(x) = \frac{B_i}{(\frac{d}{2}+x)^3} + \frac{B_i}{(\frac{d}{2}-x)^3}$. The space within $x_{min}$ from the surface is modeled as a parallel plate capacitor with capacitance $C_H$. Here we take $C_H$=400 µF/cm$^2$ which is much larger than the usual Helmholtz capacitance because of the image charge contribution[49]. Using the boundary conditions: (1) electrostatic potential in the bulk is zero $\psi(\infty) = 0$; (2) in the pore center $\frac{d\psi}{dx}\big|_{x=0} = 0$; (3) the total charge on the confining plate is zero; and (4) the confining plate has equal potential on the inner surface and the outer surface, the unique solution of the electrostatic potential $\psi(x)$ can be determined.

**Acknowledgements**

This work is supported by National Science Foundation under Grant DMR-0906547.

**Author Contributions**

Y.Z.X prepared the activated carbon material; Z.X.L and Y.C.L conducted the NMR experiments and analyzed the data; Z.X.L carried out the numerical calculation; Y.W and A.K supervised the project; Z.X.L and Y.W wrote the manuscript.

**Competing Financial Interests Statement**

The authors declare no competing financial interest.

**Figure Legends**

**Figure 1 | Ion concentrations $c/c_0$ in non-charged P-40 carbon nanopores.** (**a**) Static $^{19}$F, $^{1}$H and $^{23}$Na NMR spectra of 30 μL 1 mol/kg NaBF$_4$ electrolyte in 20 mg P-40. The peak on the left is chosen as the reference (0 ppm) and the right peak is centered at -7 ppm for all three nuclei due to NICS. (**b**) Cation and anion concentrations of NaF, NaNO$_3$, and NaBF$_4$ electrolytes inside P-40 nanopores. NaBF$_4$ electrolyte in a larger pore size sample P-92 is also shown for comparison. (**c**) Na$^+$ concentration inside nanopores for different sodium salt electrolytes plotted in the sequence of the anionic Hofmeister series.

**Figure 2 | Numerical calculation using generalized PB equation.** (**a**) Two conducting metal plates are immersed in 1M 1:1 electrolyte to simulate slit-shaped carbon nanopore with pore size $d$. The pore center is set as $x=0$. $x_{min}$ is defined as the closest distance that the ions can approach the surface. (**b**) Ion distribution in 1 nm pores for $B_+ = 46 \times 10^{-50}$ Jm$^3$, $B_- = -58 \times 10^{-50}$ Jm$^3$. (**c**) Average ion concentration in nanopores versus pore size. (**d**) Average ion concentration in nanopores versus the parameter $B_-$. The parameter $B_+ = 46 \times 10^{-50}$ is fixed.

**Figure 3 | Ion concentrations in charged P-40 nanopores versus charging voltage.** (**a**) Illustration of the device built into an NMR probe for controlling P-40 surface charging. The device is comprised of two P-40 electrodes immersed in electrolyte and separated by a glass fiber (similar to a supercapacitor). Voltage is applied between the two electrodes such that one electrode is positively charged and the other one is negatively charged. One electrode is covered by copper foil to enable single-electrode NMR measurements. (**b**) Nanoconfined ion concentrations for 1 mol/kg NaBF$_4$ electrolyte in P-40 versus charging voltage. (**c**) Comparison of Na$^+$ concentrations $c(V)/c(0)$ between NaBF$_4$ and NaNO$_3$ electrolytes versus charging voltage.

**Figure 1.**

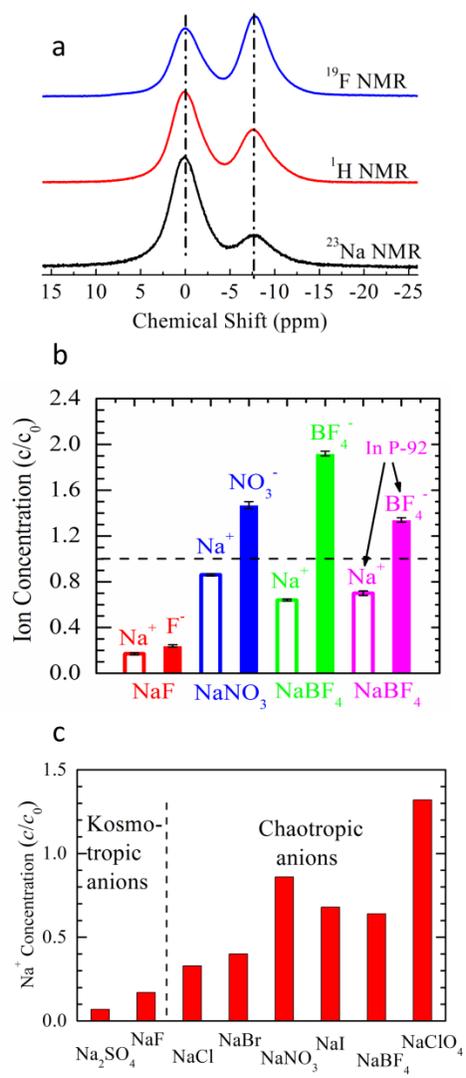

**Figure 2.**

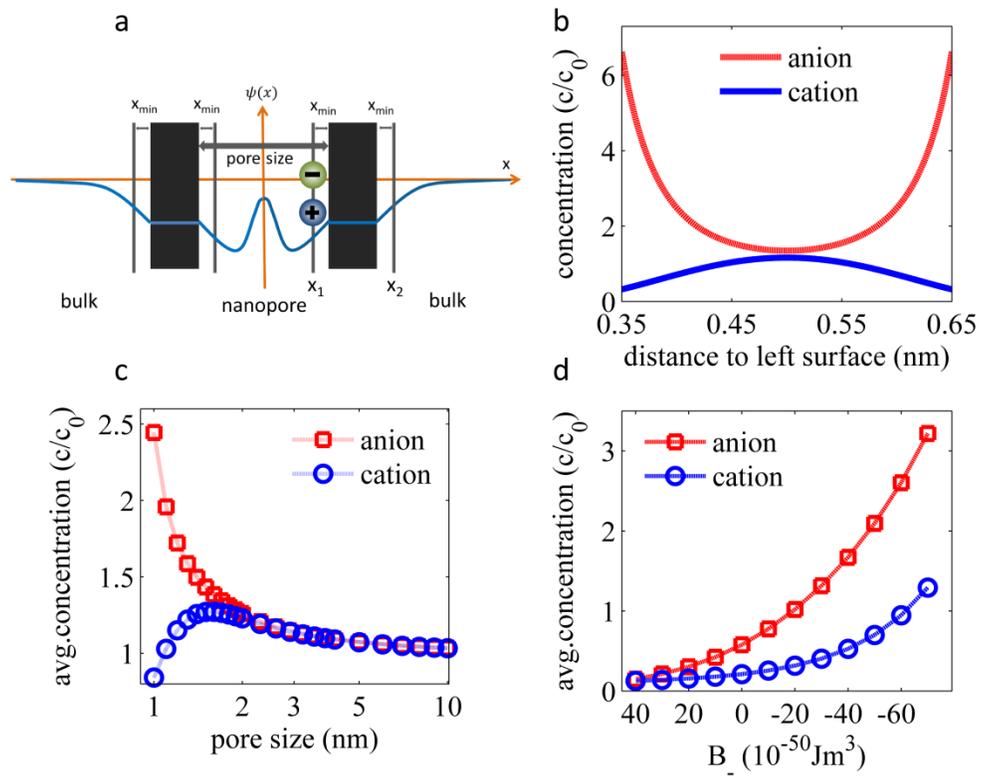

**Figure 3.**

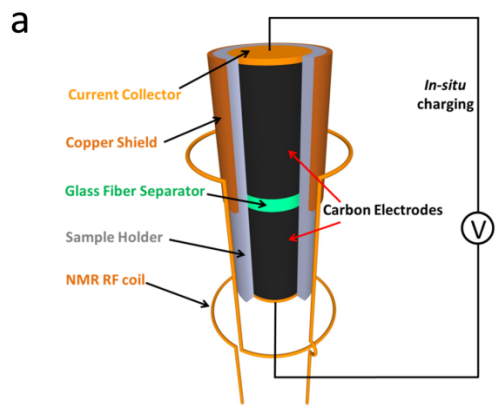

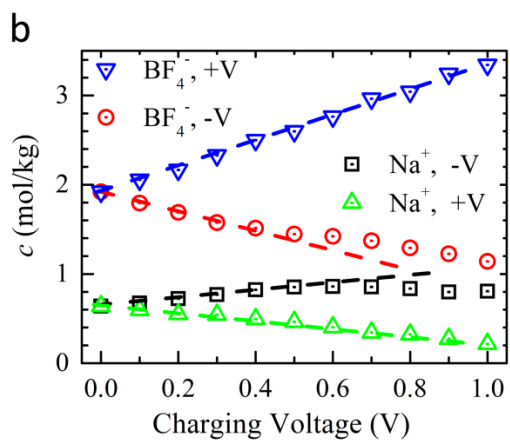

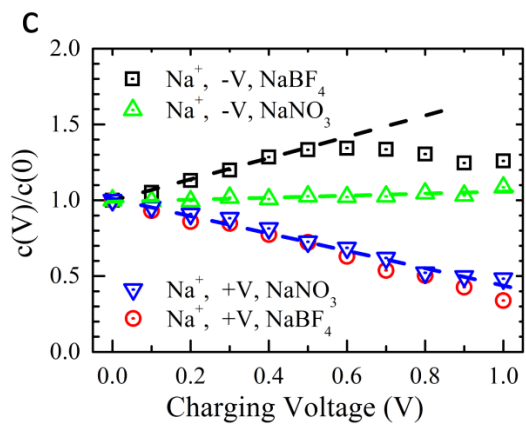